\documentclass{new_tlp}

\newif\ifarxiv
\newif\iftplp
\newif\iftplpapponly

\arxivtrue

\ifarxiv \usepackage{hyperref} \fi
\usepackage{amsfonts,latexsym,amssymb,color,amsmath}

\usepackage{comment}
\usepackage{stmaryrd}

\ifarxiv \hypersetup{breaklinks} \fi

\newcommand{\mycomment}[1]{}

\newcommand{\expk}[1]{\textup{exp}_{#1}}

\newcommand{\hide}[1]{}

\newcommand{\mo}[1]{\llbracket#1\rrbracket}

\newcommand{\mwrs}[3]{\llbracket#1\rrbracket_{#3}(#2)}

\newcommand{\aleq}[1][]{\sqsubseteq_{#1}}

\newtheorem{lemma}{Lemma}
\newtheorem{statement}{Statement}
\newtheorem{definition}{Definition}
\newtheorem{theorem}{Theorem}
\newtheorem{example}{Example}
\newtheorem{remark}{Remark}

\begin{document}

\title[The Expressive Power of Higher-Order Datalog]{The Expressive Power of Higher-Order Datalog}

\author[A. Charalambidis, Ch. Nomikos and P. Rondogiannis]
             {ANGELOS CHARALAMBIDIS\\
                Institute of Informatics and Telecommunications, NCSR ``Demokritos'', Greece \\
                \email{acharal@iit.demokritos.gr}
                \and CHRISTOS NOMIKOS\\
                Dept of Computer Science and Engineering, University of Ioannina, Greece\\
                \email{cnomikos@cs.uoi.gr}
                \and PANOS RONDOGIANNIS\\
                Dept of Informatics and Telecommunications, National and Kapodistrian University of Athens, Greece \\
            \email{prondo@di.uoa.gr}}

\pagerange{\pageref{firstpage}--\pageref{lastpage}}
\volume{\textbf{10} (3):}
\jdate{March 2002}
\setcounter{page}{1}
\pubyear{2002}

\iftplpapponly
   
\appendix

\section{The Expressive Power of First-Order Datalog}\label{PTIME}
In this appendix we present a proof of the well-known theorem~\cite{Pap85,Gra92,Var82,Imm86,Lei89}
that Datalog captures $\mathsf{PTIME}$ (under the assumption that the input strings are encoded,
as already discussed, through the {\tt input} relation). Our proof builds on that
of~\cite{Pap85} and~\cite{DEGV01}, but gives more technical details.
\begingroup
\def\thetheorem{\ref{Datalog_captures_PTIME}}
\begin{theorem}
The set of first-order Datalog programs captures $\mathsf{PTIME}$.
\end{theorem}
\addtocounter{theorem}{-1}
\endgroup

\begin{proof}
The proof of the above theorem consists of the proofs of the following two statements:
\begin{statement}\label{statement1}
Every language $L$ decided by a Datalog program $\mathsf{P}$, can also be
decided by a Turing machine in time $O(n^q)$, where $n$ is the length of the input string
and $q$ is a constant that depends only on $\mathsf{P}$.
\end{statement}

\begin{statement}\label{statement2}
Every language $L$ decided by a Turing machine in time $O(n^q)$, where
$n$ is the length of its input, can be decided by a Datalog program $\mathsf{P}$.
\end{statement}

\begin{proof*}[Proof of Statement~\ref{statement1}]
Assume that the maximum number of atoms that appear in
any rule in $\mathsf{P}$ is equal to $l$, the total number of constants that
appear in $\mathsf{P}$ is equal to $c$ (including {\tt a}, {\tt b}, {\tt empty}, {\tt end}, but
excluding the $n$ natural numbers that appear in the relation {\tt input}),
the total number of rules in $\mathsf{P}$ is
equal to $r$, the total number of predicates is equal to $p$, and the maximum arity
of a predicate that appears in $\mathsf{P}$ is $t$.
We present a multi-tape Turing machine which decides the language $L$ in time $O(n^q)$,
where $n$ is the length of the input string,
for some $q$ that depends only on the above characteristics of $\mathsf{P}$.


The Turing machine, with input $w$, starts by constructing the set of facts ${\cal D}_w$ that
represent $w$ in the Datalog program (i.e., the relation {\tt input}), which are stored on a separate tape.
Each number that appears as an argument in the relation {\tt input} is written in binary using
$O(\log n)$ bits. The construction of ${\cal D}_w$ requires $O(n \cdot \log n)$ time.

Next, the Turing machine executes the usual bottom-up procedure for computing
the least fixed-point
of a Datalog program through the iterations of the $T_{\mathsf{P}}$ operator. Intuitively,
it starts by assigning the empty relation to all predicates in $\mathsf{P}$ and at each iteration it examines
each clause of $\mathsf{P}$ and determines if it can generate any new tuples.
The relations assigned to predicates in $\mathsf{P}$ are stored each on a separate tape.
Observe that there are at most $p\cdot(n+c)^t$ tuples in the minimum Herbrand model $M_{\mathsf{P}}$ of $\mathsf{P}$
(in the extreme case where all the predicates have the same maximum arity $t$ and all possible
tuples for all possible predicates belong to the minimum model). Therefore, the bottom-up procedure
will terminate after at most $p\cdot(n+c)^t$ iterations, since at each iteration
at least one tuple must be produced. Each such iteration of the bottom-up computation takes polynomial time
with respect to $n$:
\setlength\leftmargin    {\leftmargini}

\begin{itemize}
\item For every rule, the machine instantiates all the variables using the $(n+c)$
      available constants. The number of different such instantiations
      of a rule is bounded by $(n+c)^{l\cdot t}$.

\item For each such instantiation it examines if the atoms in the body of the rule have
      already been produced in a previous step of the computation. Searching through the
      list of the already produced atoms for a specific predicate takes time
      $O(t \cdot \log n \cdot(n+c)^t)$ in the worst case
      (since the maximum number of atoms that such a list may contain is $(n+c)^t$
      and the length of each atom is $O(t \cdot \log n)$). Doing this for
      all atoms in the rule body, requires time $O(l \cdot t \cdot \log n \cdot(n+c)^t)$.
      If all atoms in the (instantiated) body of the rule are found in the corresponding lists,
      then we search the head of the rule in the list that corresponds to its predicate; if
      it is not found, then it is appended at the end of the list.
      This search and update requires time $O(t \cdot \log n \cdot(n+c)^t)$.

\item Doing the above operation for all the rules of the program requires time
      $O(r \cdot l \cdot t \cdot \log n \cdot(n+c)^{(l+1) \cdot t})$.
\end{itemize}
From the above we get that in order to produce the minimum Herbrand model $M_{\mathsf{P}}$
of $\mathsf{P}$, we need time $O(n \cdot \log n + p \cdot r \cdot l \cdot t \cdot \log n \cdot(n+c)^{(l+2) \cdot t})$.
Since $p, r,l,t$ and $c$ are constants that depend only on $\mathsf{P}$ and do not depend on $n$,
the running time of the Turing machine is $O(n^q)$ for $q = (l+2) \cdot t$.

The Turing machine returns $yes$ if and only if {\tt accept} is true in the minimum Herbrand
model $M_{\mathsf{P}}$.
\end{proof*}

\begin{proof*}[Proof of Statement~\ref{statement2}]
In order to establish the second statement, we need to define a simulator of the Turing
machine in Datalog.
Assume that $M$ decides $L$ in time $O(n^q)$. Then there exists an integer constant $d$,
such that for every input $w$ of length $n \geq 2$, $M$ terminates after at most $n^d-1$ steps.
The Datalog program that is presented below produces the correct answer for all
inputs of length at least 2 by simulating $n^d-1$ steps of the Turing machine $M$.
For the special cases of strings of length 0 or 1 that belong to $L$, the correct answer is produced
directly by appropriate rules
(notice that for $n=1$, the value of $n^d-1$ is 0, regardless of the choice of $d$).

We start by defining predicates {\tt base\_zero} (which is true of 0,
ie. of the first argument of the first tuple in the {\tt input} relation), {\tt base\_last} (which
is true of $n-1$, ie., of the first argument of the {\em last} tuple in the {\tt input}
relation), {\tt base\_succ} (which, given a number $k$, $0\leq k < n-1$, returns $k+1$), and
{\tt base\_pred} (which given $k+1$ returns $k$):
\[
\begin{array}{lll}
  \mbox{\tt base\_zero 0}.      &            & \\         
  \mbox{\tt base\_last I}       & \leftarrow & \mbox{\tt (input I X end).}\\
  \mbox{\tt base\_succ I J}     & \leftarrow & \mbox{\tt (input I X J),(input J,A,K).}\\
  \mbox{\tt base\_pred I J}     & \leftarrow & \mbox{\tt (base\_succ J I).}
\end{array}
\]

Given the above predicates, we can simulate counting from 0 up to $n-1$. We extend
the range of the numbers we can support up to $n^d-1$ for any fixed $d$ by using
$d$ distinct arguments in predicates; we view these $d$ arguments more conveniently as
$d$-tuples, ie., we use the notation $\bar{\tt X}$ to represent the sequence of $d$
arguments ${\tt X}_1,\ldots,{\tt X}_d$. We define the predicates ${\tt tuple\_zero}$,
${\tt tuple\_last}$ and ${\tt tuple\_base\_last}$ that act on such $d$-tuples and represent
the numbers 0, $n^d-1$ and $n-1$ respectively.
{\tt
\begin{center}
\begin{tabular}{lll}
tuple\_zero  $\overline{\tt X}$ & $\leftarrow$ & (base\_zero X$_1$),$\ldots$,(base\_zero X$_d$).\\
tuple\_last  $\overline{\tt X}$ & $\leftarrow$ & (base\_last X$_1$),$\ldots$,(base\_last X$_d$).\\
tuple\_base\_last  $\overline{\tt X}$ & $\leftarrow$ & (base\_zero X$_1$),$\ldots$,(base\_zero X$_{d-1}$),\\
                                      &              & (base\_last X$_d$).
\end{tabular}
\end{center}
}
To define ${\tt tuple\_succ}$ we need $d$ clauses that have as arguments two tuples having $d$ elements each:
{\tt
\begin{center}
\begin{tabular}{lll}
tuple\_succ $\overline{\tt X}$ $\overline{\tt Y}$ & $\leftarrow$ & ({\tt X}$_1$ $\approx$ {\tt Y}$_1$),...,({\tt X}$_{d-1}$ $\approx$ {\tt Y}$_{d-1}$),\\
                                                  &              &(base\_succ X$_d$ Y$_d$).\\
tuple\_succ $\overline{\tt X}$ $\overline{\tt Y}$ & $\leftarrow$ & ({\tt X}$_1$ $\approx$ {\tt Y}$_1$),...,({\tt X}$_{d-2}$ $\approx$ {\tt Y}$_{d-2}$),\\
                                                  &              & (base\_succ X$_{d-1}$ Y$_{d-1}$),\\
                                                  &              & (base\_last X$_d$),\\
                                                  &              & (base\_zero Y$_d$).\\
                                                  &    $\cdots$  &\\
tuple\_succ $\overline{\tt X}$ $\overline{\tt Y}$ & $\leftarrow$ & (base\_succ X$_{1}$ Y$_{1}$),\\
                                                  &              & (base\_last X$_2$),...,(base\_last X$_d$),\\
                                                  &              & (base\_zero Y$_2$),...,(base\_zero Y$_d$).
\end{tabular}
\end{center}
}
Now we can easily define {\tt tuple\_pred} as follows:
{\tt
\begin{center}
\begin{tabular}{lll}
tuple\_pred $\overline{\tt X}$ $\overline{\tt Y}$  & $\leftarrow$ & tuple\_succ $\overline{\tt Y}$ $\overline{\tt X}$.
\end{tabular}
\end{center}
}
The {\tt less\_than} relation over the numbers we consider, is defined as follows:
{\tt
\begin{center}
\begin{tabular}{lll}
less\_than $\overline{\tt X}$ $\overline{\tt Y}$ & $\leftarrow$ & tuple\_succ $\overline{\tt X}$ $\overline{\tt Y}$.\\
less\_than $\overline{\tt X}$ $\overline{\tt Y}$ & $\leftarrow$ & (tuple\_succ $\overline{\tt X}$  $\overline{\tt Z}$),(less\_than $\overline{\tt Z}$ $\overline{\tt Y}$).
\end{tabular}
\end{center}
}
We can also define {\tt tuple\_non\_zero}, namely the predicate that succeeds if its argument is not
equal to zero:
{\tt
\begin{center}
\begin{tabular}{lll}
tuple\_non\_zero $\overline{\tt X}$ & $\leftarrow$ & (tuple\_zero $\overline{\tt Z}$),(less\_than $\overline{\tt Z}$ $\overline{\tt X}$).
\end{tabular}
\end{center}
}
We now define predicates ${\tt symbol}_\sigma$, ${\tt state}_s$ and ${\tt cursor}$,
for every $\sigma \in \Sigma$ and for every state $s$ of the Turing machine we are
simulating. Intuitively, ${\tt symbol}_{\sigma} \ \overline{\tt T} \ \overline{\tt X}$
succeeds if the tape has symbol $\sigma$ in position $\overline{\tt X}$ of the tape
at time-step $\overline{\tt T}$,  ${\tt state}_s \ \overline{\tt T}$ succeeds if the
machine is in state $s$ at step $\overline{\tt T}$ and ${\tt cursor}\ \overline{\tt T}\ \overline{\tt X}$
succeeds if the head of the machine points at position $\overline{\tt X}$ at step $\overline{\tt T}$.
Since symbols and states are finite there will be a finite number of clauses defining
the above predicates. We assume that the Turing machine never attempts to go to the left
of its leftmost symbol. Moreover, we assume that in the beginning of its operation,
the first $n$ squares of the tape hold the input, the rest of the squares hold the
empty character ``\textvisiblespace'' and the machine starts operating from its initial
state denoted by $s_0$. If the Turing machine accepts the input then it goes into the
special state called {\tt yes} and stays there forever.

The initialization of the Turing machine is performed by the following clauses:
{\tt
\begin{center}
\begin{tabular}{lll}
symbol$_\sigma$ $\overline{\tt T}$ $\overline{\tt X}$  & $\leftarrow$ & (tuple\_zero $\overline{\tt T}$),\\
&        & (base\_zero X$_1$),$\ldots$,(base\_zero X$_{d-1}$),\\
&        & (input X$_d$ $\sigma$ W).\\
symbol$_{\textvisiblespace}$ $\overline{\tt T}$ $\overline{\tt X}$  & $\leftarrow$ & (tuple\_zero $\overline{\tt T}$),\\
&        & (tuple\_base\_last $\overline{\tt Y}$),(less\_than $\overline{\tt Y}$ $\overline{\tt X}$).\\

state$_{s_0}$ $\overline{\tt T}$ & $\leftarrow$ & (tuple\_zero $\overline{\tt T}$).\\
cursor $\overline{\tt T}$ $\overline{\tt X}$ & $\leftarrow$ & (tuple\_zero $\overline{\tt T}$),(tuple\_zero $\overline{\tt X}$).
\end{tabular}
\end{center}
}
For each transition rule we generate a set of clauses. We start with
the rule ``if the head is in symbol $\sigma$ and in state $s$ then
write symbol $\sigma'$ and go to state $s'$'', which is translated as follows:
{\tt
\begin{center}
\begin{tabular}{lll}
symbol$_{\sigma'}$ $\overline{\tt T'}$ $\overline{\tt X}$& $\leftarrow$ &(tuple\_succ $\overline{\tt T}$ $\overline{\tt T'}$),(state$_s$ $\overline{\tt T}$),\\
&              & (cursor $\overline{\tt T}$ $\overline{\tt X}$),(symbol$_\sigma$ $\overline{\tt T}$ $\overline{\tt X}$).\\
state$_{s'}$ $\overline{\tt T'}$ & $\leftarrow$ & (tuple\_succ $\overline{\tt T}$ $\overline{\tt T'}$),(state$_s$ $\overline{\tt T}$),\\
&              &(cursor $\overline{\tt T}$ $\overline{\tt X}$),(symbol$_\sigma$ $\overline{\tt T}$ $\overline{\tt X}$). \\
cursor $\overline{\tt T'}$ $\overline{\tt X}$ & $\leftarrow$  & (tuple\_succ $\overline{\tt T}$ $\overline{\tt T'}$),(state$_s$ $\overline{\tt T}$),\\
&              &(cursor $\overline{\tt T}$ $\overline{\tt X}$),(symbol$_\sigma$ $\overline{\tt T}$ $\overline{\tt X}$).
\end{tabular}
\end{center}
}
We continue with the transition: ``if the head is in symbol $\sigma$ and in state $s$, then go to state $s'$ and move the
head right'', which generates the following:
{\tt
\begin{center}
\begin{tabular}{lll}
symbol$_{\sigma}$ $\overline{\tt T'}$ $\overline{\tt X}$& $\leftarrow$ &(tuple\_succ $\overline{\tt T}$ $\overline{\tt T'}$),(state$_s$ $\overline{\tt T}$),\\
&              & (cursor $\overline{\tt T}$ $\overline{\tt X}$),(symbol$_\sigma$ $\overline{\tt T}$ $\overline{\tt X}$).\\
state$_{s'}$ $\overline{\tt T'}$ & $\leftarrow$ & (tuple\_succ $\overline{\tt T}$ $\overline{\tt T'}$),(state$_s$ $\overline{\tt T}$),\\
&              &(cursor $\overline{\tt T}$ $\overline{\tt X}$),(symbol$_\sigma$ $\overline{\tt T}$ $\overline{\tt X}$). \\
cursor $\overline{\tt T'}$ $\overline{\tt X'}$ & $\leftarrow$  & (tuple\_succ $\overline{\tt T}$ $\overline{\tt T'}$),(state$_s$ $\overline{\tt T}$),\\
&              &(cursor $\overline{\tt T}$ $\overline{\tt X}$),(symbol$_\sigma$ $\overline{\tt T}$ $\overline{\tt X}$),(tuple\_succ $\overline{\tt X}$ $\overline{\tt X'}$).
\end{tabular}
\end{center}
}
We also have the transition: ``if the head is in symbol $\sigma$ and in state $s$ then
go to state $s'$ and move the head left'', which generates the following:
{\tt
\begin{center}
\begin{tabular}{lll}
symbol$_{\sigma}$ $\overline{\tt T'}$ $\overline{\tt X}$& $\leftarrow$ &(tuple\_succ $\overline{\tt T}$ $\overline{\tt T'}$),(state$_s$ $\overline{\tt T}$),\\
&              & (cursor $\overline{\tt T}$ $\overline{\tt X}$),(symbol$_\sigma$ $\overline{\tt T}$ $\overline{\tt X}$).\\
state$_{s'}$ $\overline{\tt T'}$ & $\leftarrow$ & (tuple\_succ $\overline{\tt T}$ $\overline{\tt T'}$),(state$_s$ $\overline{\tt T}$),\\
&              &(cursor $\overline{\tt T}$ $\overline{\tt X}$),(symbol$_\sigma$ $\overline{\tt T}$ $\overline{\tt X}$). \\
cursor $\overline{\tt T'}$ $\overline{\tt X'}$ & $\leftarrow$  & (tuple\_succ $\overline{\tt T}$ $\overline{\tt T'}$),(state$_s$ $\overline{\tt T}$),\\
&              &(cursor $\overline{\tt T}$ $\overline{\tt X}$),(symbol$_\sigma$ $\overline{\tt T}$ $\overline{\tt X}$),(tuple\_pred $\overline{\tt X}$ $\overline{\tt X'}$).
\end{tabular}
\end{center}
}
We also need to provide ``inertia'' rules for the tape squares
that are not affected by the above rules. These squares are exactly those that have
a different position from the one pointed to by the cursor. Therefore, the following
clauses will suffice:
{\tt
\begin{center}
\begin{tabular}{lll}
symbol$_{\sigma}$ $\overline{\tt T'}$ $\overline{\tt X'}$ & $\leftarrow$  & (tuple\_succ $\overline{\tt T}$ $\overline{\tt T'}$),(cursor $\overline{\tt T}$ $\overline{\tt X}$),\\
&           &(less\_than $\overline{\tt X}$ $\overline{\tt X'}$),(symbol$_\sigma$ $\overline{\tt T}$ $\overline{\tt X'}$). \\
symbol$_{\sigma}$ $\overline{\tt T'}$ $\overline{\tt X'}$ & $\leftarrow$  & (tuple\_succ $\overline{\tt T}$ $\overline{\tt T'}$),(cursor $\overline{\tt T}$ $\overline{\tt X}$),\\
&           &(less\_than $\overline{\tt X'}$ $\overline{\tt X}$),(symbol$_\sigma$ $\overline{\tt T}$ $\overline{\tt X'}$).
\end{tabular}
\end{center}
}
Lastly, the following rule succeeds iff the Turing machine succeeds after $n^d-1$ steps.
{\tt
\begin{center}
\begin{tabular}{lll}
accept & $\leftarrow$ & (tuple\_last $\overline{\tt T}$),(state$_{\tt yes}$ $\overline{\tt T}$).
\end{tabular}
\end{center}
}
In the case of strings of length $n\leq 1$ that belong to $L$, we add appropriate rules to the
Datalog program. For example, if $a\in L$, then the following rule is included
in the Datalog program:
{\tt
\begin{center}
\begin{tabular}{lll}
accept & $\leftarrow$ & (input 0 a end).
\end{tabular}
\end{center}
}
\end{proof*} 
This completes the proof of the theorem.
\end{proof}

\pagebreak
\section{Proof of Lemma~\ref{first_direction}}\label{appendix_b}

\begingroup
\def\thelemma{\ref{first_direction}}
\begin{lemma}
Let $\mathsf{P}$ be a $k$-order Datalog program, $k\geq 2$, that decides
a language $L$. Then, there exists a Turing machine that decides the same
language in time $O(\expk{k-1}(n^q))$, where $n$ is the length of the input
string and $q$ is a constant that depends only on $\mathsf{P}$.
\end{lemma}
\addtocounter{lemma}{-1}
\endgroup
\begin{proof}
Let $\mathsf{P}$ be a $k$-order Datalog program that decides a language $L$.
Assume that the maximum length of a rule in $\mathsf{P}$ is equal to $l$,
the total number of constants that appear in $\mathsf{P}$ is equal to $c$
(including {\tt a}, {\tt b}, {\tt empty}, {\tt end}, but
excluding the $n$ natural numbers that appear in the relation {\tt input}),
the total number of rules in $\mathsf{P}$ is equal to $r$,
the total number of predicates is equal to $p$,
the total number of predicates types involved in $\mathsf{P}$ is equal to $s$,
and the maximum arity of a predicate type that is involved in $\mathsf{P}$ is $t$.
A summary of all these parameters is given in Table B1.
\begin{table}[h!]
\begin{center}
\begin{tabular}{cl}
 \hline\hline
 {\em Symbol}   & {\em Characteristic of $\mathsf{P}$}\\ \hline\hline
     $l$        & maximum length of a rule\\
     $c$        & number of constants\\
     $r$        & number of rules\\
     $p$        & number of predicates\\
     $s$        & number of predicate types\\
     $t$        & maximum arity of a predicate type\\ \hline
\end{tabular}
\caption{Characteristics of $\mathsf{P}$ used in our analysis.}
\end{center}
\end{table}
We present a multi-tape Turing machine which decides the language $L$ in time $O(\expk{k-1}(n^q))$,
where $n$ is the length of the input string, for some $q$ that depends only
on the above characteristics of $\mathsf{P}$.

The Turing machine, with input $w$, starts by constructing the set of facts ${\cal D}_w$ that
represent $w$ in the Datalog program (i.e., the relation {\tt input}), which are stored on a separate tape.
Each number that appears as an argument in the relation {\tt input} is written in binary using
$O(\log n)$ bits. It also writes on a separate tape all the elements of the set $\mo{\iota}$
(that is, the $c$ constants that occur in $\mathsf{P}$ and the $n$ numbers that appear in the input relation).
This requires $O(n \cdot \log n)$ time.

Subsequently, the Turing machine performs two major phases: (i) it produces all the monotonic
relations that are needed for the bottom-up execution of the program, and (ii) it performs
instantiations of the rules using these monotonic relations as-well-as individual constants,
computing in this way, in a bottom-up manner, the minimum Herbrand model of ${\mathsf P}$.
The complexity of these two major phases is analyzed in detail below.

\paragraph*{Complexity of producing the monotonic relations.} 
The Turing machine constructs the set $\mo{\rho}$,
for every predicate type $\mo{\rho}$ of order at most $k-1$, which is involved in $\mathsf{P}$.
The elements of $\mo{\rho}$ are monotonic functions, which are represented
by their corresponding relations.
Predicate types are considered in increasing order, and for each type $\mo{\rho}$
the set $\mo{\rho}$ is stored on a separate tape. These sets will be used later by the Turing machine,
each time that it needs to instantiate predicate variables that occur in the rules of  $\mathsf{P}$.

Before we present in more details the above construction and analyze its time complexity,
we need to calculate upper bounds for the number of elements in $\mo{\rho}$ and
for the length of their representation.
We prove that, for every $j$-order predicate $\rho$,
the number of elements in $\mo{\rho}$ is at most $\expk{j}(t^{j-1} \cdot (n+c)^t)$
and each of them can be represented using
$O(\log n \cdot \expk{j-1}(j \cdot t^{j} \cdot (n+c)^t))$
symbols. These two statements can be proved simultaneously by induction on $j$, as shown below.

For the basis of the induction, consider a first order predicate type $\rho$ of arity $m \leq t$.
Each element in $\mo{\rho}$ corresponds to a set of tuples, where each tuple consists of $m$
constants. Since there are $(n+c)$ different constants in $\mathsf{P} \cup {\cal D}_w$,
there are $2^{(n+c)^m} \leq 2^{(n+c)^t} = \expk{1}(t^{0} \cdot (n+c)^t)$
elements in $\mo{\rho}$.
Moreover, every element in $\mo{\rho}$ contains at most $(n+c)^t$ tuples, each tuple consists
of at most $t$ constants and each constant can be represented using $O(\log n)$ symbols.
Thus the length of the representation of every element in $\mo{\rho}$ is $O(\log n \cdot t \cdot (n+c)^t)
= O(\log n \cdot \expk{0}(1 \cdot t^1 \cdot (n+c)^t))$. Thus, the statement holds for $j=1$.

For the induction step, assume that $j>1$ and that our statement holds for all $i<j$.
Consider a $j$-order predicate type $\rho = \rho_1 \rightarrow \cdots \rightarrow \rho_m \rightarrow o$
of arity $m \leq t$.
Each element in $\mo{\rho}$ corresponds to a subset of $\mo{\rho_1} \times \cdots \times \mo{\rho_m}$.
For every $\nu$, $1 \leq \nu \leq m$, $\rho_\nu$ is either equal to $\iota$,
or is an $i$-order predicate type, with $i<j$.
In the former case, $\mo{\rho_\nu}$ contains $(n+c)$ elements; in the latter case
$\mo{\rho_\nu}$ contains at most $\expk{i}(t^{i-1} \cdot (n+c)^t)$ elements, by the induction hypothesis.
In both cases $\mo{\rho_\nu}$ contains at most $\expk{j-1}(t^{j-2} \cdot (n+c)^t)$ elements.
Using some properties of the function $\expk{}$, we get that the number of elements in $\mo{\rho}$ is bounded by
$2^{(\expk{j-1}(t^{j-2} \cdot (n+c)^t))^m} \leq
2^{\expk{j-1}(m \cdot t^{j-2} \cdot (n+c)^t)}
\leq 2^{\expk{j-1}(t^{j-1} \cdot (n+c)^t)}
= \expk{j}(t^{j-1} \cdot (n+c)^t)$.
Moreover, every element in $\mo{\rho}$ corresponds to a relation that contains at most
$(\expk{j-1}(t^{j-2} \cdot (n+c)^t))^t \leq \expk{j-1}(t^{j-1} \cdot (n+c)^t)$
tuples; each one of these tuples consists of at most $t$ elements and, by the induction hypothesis,
each element can be represented using
$O(\log n \cdot \expk{j-2}((j-1) \cdot t^{j-1} \cdot (n+c)^t))$ symbols.
By the properties of the function $\expk{}$ it follows that
$t \cdot \expk{j-2}((j-1) \cdot t^{j-1} \cdot (n+c)^t)
\leq \expk{j-2}((j-1) \cdot t^j \cdot (n+c)^t)
\leq \expk{j-1}((j-1) \cdot t^j \cdot (n+c)^t)$ and
$\expk{j-1}(t^{j-1} \cdot (n+c)^t) \cdot \expk{j-1}((j-1) \cdot t^j \cdot (n+c)^t)
\leq \expk{j-1}(j \cdot t^j \cdot (n+c)^t)$. Thus,
the length of the representation of a $j$-order relation is
$O(\log n \cdot \expk{j-1}(j \cdot t^{j} \cdot (n+c)^t))$.

Since $c,t$ and $j$ are constants that do not depend on $n$,
it follows that for every $j$-order predicate $\rho$,
the number of elements in $\mo{\rho}$
is $O(\expk{j}(n^{t+1}))$ and each of these elements
can be represented using $O(\expk{j-1}(n^{t+1}))$ symbols.

In order to create a list with all the elements of type $\mo{\rho}$ for a $j$-order type
$\rho = \rho_1 \rightarrow \cdots \rightarrow \rho_m \rightarrow o$ of arity $m \leq t$,
the Turing machine first constructs the cartesian product
$S = \mo{\rho_1} \times \cdots \times \mo{\rho_m}$.
Since $\rho_1, \cdots, \rho_m$ have order at most $j-1$,
the sets $\mo{\rho_1}, \cdots, \mo{\rho_m}$ have already been constructed in previous steps of the Turing machine.
The construction of $S$ requires time linear to the length of its representation, provided that the sets
$\mo{\rho_1}, \cdots, \mo{\rho_m}$ are stored on separate tapes.
Since $\rho$ may have arguments of the same type, this may require to create at most $m-1$ copies
of such sets.
Notice that the sets $S, \mo{\rho_1}, \cdots, \mo{\rho_m}$ are actually $j$-order relations,
and therefore their representations have length $O(\expk{j-1}(n^{t+1}))$.
We conclude that the time
required to create $S$ is $O(\expk{j-1}(n^{t+1}))$ (since $m$ is a constant that does not depend on $n$).

The set $\mo{\rho}$ contains the elements in the powerset $2^S$ of $S$ which represent monotonic functions.
The set $2^S$ can be constructed in time linear to the length of its representation. Since
$2^S$ is a $(j+1)$-order relation, this length is $O(\expk{j}(n^{t+1}))$.
Thus, $2^S$ can be constructed in time $O(\expk{j}(n^{t+1}))$.
Now, $\mo{\rho}$ can be obtained from $2^S$, by removing elements that represent
non-monotonic functions.
In order to decide whether a relation in $2^S$ belongs to $\mo{\rho}$, it suffices to consider
each pair of elements in $S$ and verify that, for this pair, the monotonicity property is not violated.
This verification requires time linear to the length of the relation, for each pair of elements in $S$.
Thus, the time required to check whether a relation represents a monotonic function is
$O((\expk{j-1}(n^{t+1}))^{2t}
\cdot \expk{j-1}(n^{t+1}))$.
The number of relations in $2^S$ is $O(\expk{j}(n^{t+1}))$.
Using the properties of the function $\expk{}$, we get that
$(\expk{j-1}(n^{t+1}))^{2t} \cdot \expk{j-1}(n^{t+1}) \cdot \expk{j}(n^{t+1})
\leq \expk{j}(2t \cdot n^{t+1}) \cdot \expk{j}(n^{t+1}) \cdot \expk{j}(n^{t+1})
\leq \expk{j}((2t+2) \cdot n^{t+1}) \leq \expk{j}(n^{t+2})$.
Thus, the removal of relations that do not correspond to monotonic functions
requires time $O(\expk{j}(n^{t+2}))$.

By adding the times required to construct $S$ and $2^S$, and remove non-relevant relations,
we conclude that the time to create a list with all the elements in $\mo{\rho}$ is
$O(\expk{j-1}(n^{t+1}))+O(\expk{j}(n^{t+1}))+O(\expk{j}(n^{t+2})) = O(\expk{j}(n^{t+2}))$.

The above process is executed for each of the $s$ predicate types of order at most $k-1$ involved in $\mathsf{P}$.
Since $s$ is a constant that does not depend on $n$, the time required for the construction
of all the relations of each predicate type is $O(\expk{k-1}(n^{t+2}))$.

\paragraph*{Complexity of performing the bottom-up computation.}
Next, the Turing machine essentially computes the successive approximations
to the minimum Herbrand model $M_{\mathsf{P}}$ of $\mathsf{P}$, by iterative application of the
$T_{\mathsf{P}}$ operator (as described at the end of Section~\ref{syntax_and_semantics}).
For each predicate constant defined in $\mathsf{P}$, the relation that represents its meaning
is written on a separate tape; initially all these relations are empty.
At each iteration of the $T_{\mathsf{P}}$ operator, new tuples may be added to the
meaning of predicate constants.
In order to calculate one
iteration of $T_{\mathsf{P}}$, the Turing machine considers each clause in $\mathsf{P}$ and examines
what new tuples it can produce.
More specifically, given a rule, it replaces every
individual variable that appears in the rule by a constant symbol and every predicate
variable with a monotonic relation of the same type as the variable; moreover, it
replaces every predicate constant, say $\mathsf{q}$, that appears in the body of
the clause with the relation that has been computed for $\mathsf{q}$
during the previous iterations of $T_{\mathsf{P}}$.
It then checks if the body of the
instantiated clause evaluates to true: this is performed by essentially checking if
elements belong to sets.
If an instantiation of a clause body evaluates to true,
the instantiated head is added to the meaning of the head predicate.

Observe that there are at most $p \cdot (\expk{k-1}(t^{k-2} \cdot (n+c)^t))^t$ tuples in the
minimum Herbrand model $M_{\mathsf{P}}$ of $\mathsf{P}$
(in the extreme case, all the predicates have order $k$, the same maximum arity $t$ and all possible
tuples for all possible predicates belong to the minimum model).
Therefore, the bottom-up procedure
will terminate after at most $p \cdot (\expk{k-1}(t^{k-2} \cdot (n+c)^t))^t$ iterations, since at each iteration
at least one tuple must be produced. Since
$(\expk{k-1}(t^{k-2} \cdot (n+c)^t))^t \leq \expk{k-1}(t^{k-1} \cdot (n+c)^t)$
and $k,p,t$ are constants that do not depend on $n$, the number of iterations is
$O(\expk{k-1}(n^{t+1}))$.

We calculate a bound of the time
that is required for each one of the above iterations:

\setlength\leftmargin    {\leftmargini}
\begin{itemize}
\item For every rule in the program, the Turing machine instantiates each individual variable
      in the rule using elements in $\mo{\iota}$. Moreover, it instantiates each
      predicate variables of type $\rho$ with relations representing monotonic functions in $\mo{\rho}$
      (recall that these sets have been constructed in the first phase of the execution of the Turing machine).
      Finally, it replaces every predicate constant
      in the body of the rule with the relation that has already been computed for it
      during the previous iterations of $T_{\mathsf{P}}$.
      By the syntactic rules of Higher-Order Datalog programs, predicate variables may have order at most $k-1$.
      Thus, the number of different such instantiations
      of a rule is bounded by $(\expk{k-1}(t^{k-2} \cdot (n+c)^t))^l$. Since
      $(\expk{k-1}(t^{k-2} \cdot (n+c)^t))^l \leq \expk{k-1}(l\cdot t^{k-2} \cdot (n+c)^t)$
      and $k,l,t$ are constants that do not depend on $n$, the number of different instantiations for each rule is
      $O(\expk{k-1}(n^{t+1}))$.

\item Each rule contains a constant number of (individual or predicate) variables. Since
      all predicate variables have order at most $k-1$, each variable is replaced by at most
      $O(\expk{k-2}(n^{t+1}))$ symbols. Thus, the length of the instantiated rule is $O(\expk{k-2}(n^{t+1}))$.
      Moreover, the instantiation can be computed in time linear to its length.

\item For each such instantiation the Turing machine examines if the body of the rule evaluates to {\em true}.
      This may require at most $l$ rewritings of the instantiated body of the rule,
      each resulting after partially applying a predicate to its first argument.
      Each rewriting requires time linear to the length of the instantiated rule, that is,
      $O(\expk{k-2}(n^{t+1}))$. Since $l$ does not depend on $n$, the total time that is needed to
      examine if the body of the rule evaluates to {\em true} is $O(\expk{k-2}(n^{t+1}))$.

\item If the body of some instantiated rule evaluates to {\em true},
      then we search the head of the rule in the list that corresponds to its predicate; if
      it is not found, then it is inserted in the list.
      This search and insertion requires time linear to the length of this list,
      which is $O(\expk{k-1}(n^{t+1}))$.

\item The total time needed to repeat the above process for every instantiation of a specific rule is
      $O(\expk{k-1}(n^{t+1})) \cdot O(\expk{k-1}(n^{t+1})) = O((\expk{k-1}(n^{t+1}))^2) $.

\item Since the number of rules $r$ does not depend on $n$,
      the total time required for one iteration of the $T_{\mathsf{P}}$ operator is also
      $O((\expk{k-1}(n^{t+1}))^2)$.

\end{itemize}
From the above we get that in order to produce the minimum Herbrand model $M_{\mathsf{P}}$
of $\mathsf{P}$, we need time
$O(\expk{k-1}(n^{t+1})) \cdot O((\expk{k-1}(n^{t+1}))^2) = O((\expk{k-1}(n^{t+1}))^3)$.
By the properties of the function $\expk{}$, it is
$(\expk{k-1}(n^{t+1}))^3 \leq \expk{k-1}(3n^{t+1}) \leq \expk{k-1}(n^{t+2})$.

We conclude that the running time of the Turing machine is $O(\expk{k-1}(n^q))$ for $q = t+2$.
The Turing machine returns $yes$ if and only if {\tt accept} is true in the minimum Herbrand
model $M_{\mathsf{P}}$. This completes the proof of the lemma.
\end{proof}
\else
   \iftplp
      
\maketitle

\begin{abstract}
A classical result in descriptive complexity theory states that Datalog
expresses exactly the class of polynomially computable queries on ordered
databases~\cite{Pap85,Gra92,Var82,Imm86,Lei89}. In this paper we extend
this result to the case of higher-order Datalog. In particular, we demonstrate
that on ordered databases, for all $k\geq 2$, $k$-order Datalog captures
$(k-1)$-$\mathsf{EXPTIME}$. This result suggests that higher-order extensions
of Datalog possess superior expressive power and they are worthwhile of
further investigation both in theory and in practice. \ifarxiv This paper is under
consideration for acceptance in TPLP.\fi
\end{abstract}

\begin{keywords}
Datalog, Higher-Order Logic Programming, Descriptive Complexity Theory.
\end{keywords}

\section{Introduction}
Higher-order programming languages are widely recognized as offering a more modular
and expressive way of programming. The use of higher-order constructs in functional
programming has been a key factor for the development and success of the functional
paradigm. Functional programmers have embraced the higher-order style of programming
because they have realized that it offers significant advantages in everyday programming.
Apart from the empirical evidence of the strengths of higher-order functions, there also
exist concrete theoretical results that support this claim. For example, it has been
demonstrated~\cite{Jones} that if we restrict attention to a functional language that
is not Turing-complete, then its higher-order fragments capture broader complexity
classes than the lower-order ones.

The situation in logic programming is not so clear-cut. Logic programming languages have
traditionally been first-order, offering to programmers only certain restricted higher-order
capabilities. There have been some serious attempts to develop general-purpose higher-order
logic programming languages, most notably Hilog~\cite{CKW93-187} and $\lambda$-Prolog~\cite{MN86}.
Although these systems have not become mainstream, they have found some remarkable application domains
beyond those of traditional logic programming. For example, $\lambda$-Prolog has been used
for theorem-proving and program analysis. Moreover, Hilog's ideas have been incorporated in
the Flora-2 system which has been used for meta-programming~\cite{flora2} and data
integration~\cite{hilog-data-integration}. Also, it has recently been demonstrated that
higher-order logic programming can be used to concisely represent complicated
user-preferences in deductive databases~\cite{CRT18}, with some demonstrated applications in airline
reservation and movie-selection systems. All the above applications suggest that higher-order
logic programming can open new, fresh, and promising directions for logic programming as a whole.

In this paper we provide theoretical results that affirm the power of higher-order logic
programming. Intuitively speaking, we demonstrate that Higher-Order Datalog possesses
superior expressive power compared to classical Datalog. Our results belong to a research
stream that studies the expressive power of fragments or extensions of logic programming
languages using complexity-theoretic tools. A classical expressibility theorem in this area states
that on ordered databases
Datalog captures $\mathsf{PTIME}$~\cite{Pap85,Gra92,Var82,Imm86,Lei89}.
This is an interesting (and somewhat unexpected) result, because it suggests that a seemingly simple
language can express all polynomially computable queries. In this paper we extend this classical
result to the case of Higher-Order Datalog. More specifically:
\begin{itemize}
%

\item We demonstrate that every language decided by a $k$-order Datalog program, $k\geq 2$,
      can also be decided by a $(k-1)$-exponential-time bounded Turing machine. This result
      relies on developing a bottom-up proof procedure for $k$-order Datalog programs,
      which generalizes the familiar one for classical Datalog programs.

\item We demonstrate that  every language decided by a $(k-1)$-exponential-time bounded
      Turing machine, $k\geq 2$, can also be decided by a $k$-order Datalog program. The proof actually
      involves a simulation of the Turing machine by the Higher-Order Datalog program.
      Our simulation uses the encoding of ``big numbers'' by higher-order functions
      (relations in our case) used in~\cite{Jones}.
\end{itemize}
The above results essentially demonstrate that higher-order extensions
of Datalog possess superior expressive power than classical Datalog and
they are worthwhile of further investigation both in theory and in practice.
Additionally, the results show a striking analogy with the expressibility
results of~\cite{Jones} regarding higher-order ``read-only'' functional
programs. It may be possible that this analogy can be further exploited
to derive additional complexity-theoretic insights for interesting classes
of Higher-Order Datalog programs. This possibility is discussed in the
concluding section of the paper.

The rest of the paper is organized as follows. Section~\ref{syntax_and_semantics}
introduces the syntax and the semantics of Higher-Order Datalog. Section~\ref{complexity_basics}
presents the underlying framework for connecting logic programming with complexity theory.
Section~\ref{equivalence} demonstrates that on ordered databases, for all $k\geq 2$,
$k$-order Datalog captures $(k-1)$-$\mathsf{EXPTIME}$. Section~\ref{future_work}
gives pointers to future work. \ref{PTIME} contains a version of the proof that Datalog
captures $\mathsf{PTIME}$ on ordered databases; this is needed because our developments
are based on extending this classical result. \ref{appendix_b} contains the one direction
of our proof that $k$-order Datalog captures $(k-1)$-$\mathsf{EXPTIME}$.

\section{Higher-Order Datalog}\label{syntax_and_semantics}
\subsection{The Syntax of Higher-Order Datalog}\label{syntax}
The language {\em Higher-Order Datalog} that we consider in this paper is the function-free
subset of the higher-order logic programming language ${\cal H}$ introduced in~\cite{CHRW13}.
Higher-Order Datalog inherited from ${\cal H}$ an important syntactic restriction which ensures
that the language retains the usual least fixpoint semantics of classical logic programming.
This syntactic restriction was proposed many years ago by W. W. Wadge~\cite{Wa91a} (and also
later independently by M. Bezem~\cite{Bezem99}):

\vspace{0.16cm}
\noindent
{\bf The higher-order syntactic restriction.} {\em In the head of every rule in a program,
each argument of predicate type must be a variable, and all such variables must be distinct.}

\begin{example}
The following is a legitimate higher-order program that defines the union of two
relations {\tt P}, {\tt Q} (for the moment we use ad-hoc Prolog-like syntax):
\[
\begin{array}{l}
\mbox{\tt union(P,Q,X):-P(X).}\\
\mbox{\tt union(P,Q,X):-Q(X).}
\end{array}
\]
However, the following program does not satisfy Wadge's restriction:
\[
\begin{array}{l}
\mbox{\tt q(a).}\\
\mbox{\tt r(q).}
\end{array}
\]
because the predicate constant {\tt q} appears as an argument in the head of a rule. Similarly,
the program:
\[
\begin{array}{l}
\mbox{\tt p(Q,Q):-Q(a).}
\end{array}
\]
is problematic because the predicate variable {\tt Q} is used twice in the
head of the rule.\mathproofbox
\end{example}

We now proceed to the exact definition of the syntax of Higher-Order Datalog.
The language is based on a simple type system that supports two base types:
$o$, the boolean domain, and $\iota$, the domain of individuals (data objects). The composite
types are partitioned into two classes:  predicate (assigned to predicate symbols) and
argument (assigned to parameters of predicates).

\begin{definition}
{\em Predicate} and {\em argument} types, denoted by $\pi$
and $\rho$ respectively, are defined as follows:
\begin{align*}
\pi & := o \mid (\rho \rightarrow \pi) \\
\rho & := \iota \mid (\rho \rightarrow \pi)
\end{align*}
\end{definition}

As usual, the binary operator $\rightarrow$ is right-associative.
It can be easily seen that every predicate type $\pi$ can be written in the
form $\rho_1 \rightarrow \cdots \rightarrow \rho_n \rightarrow o$,
$n\geq 0$, and $n$ will be called the {\em arity} of the type $\pi$;
for $n=0$ we assume that $\pi=o$. We proceed by defining the
syntax of Higher-Order Datalog:
\begin{definition}
The {\em alphabet} of Higher-Order Datalog consists of the following:
\begin{itemize}
  \item Predicate constants of every predicate type $\pi$ (denoted by lowercase letters or
        words that start with lowercase letters such as $\mathsf{p,q,is\_zero,\ldots}$).
  \item Predicate variables of every predicate type $\pi$ (denoted by capital letters such as
        $\mathsf{P,Q,R,\ldots}$).
  \item Individual constants of type $\iota$ (denoted by lowercase letters or words that
        start with lowercase letters such as $\mathsf{a,b,end,\ldots}$).
  \item Individual variables of type $\iota$ (denoted by capital letters such as
      $\mathsf{X,Y,Z,\ldots}$).
  \item The inverse implication constant $\leftarrow$, the conjunction symbol $\wedge$, the left and right parentheses,
        and the equality constant $\approx$ for comparing terms of type $\iota$.
\end{itemize}
It will always be obvious from context whether a variable name that we use is a predicate or individual one;
similarly for the names of predicate and individual constants.
\end{definition}
Predicate constants correspond to the predicates that are defined in a program,
while predicate and individual variables are used as formal parameters in such predicate definitions.
The set consisting of the predicate variables and the individual variables
will be called the set of {\em argument variables}. Argument variables will
be usually denoted by $\mathsf{V}$ and its subscripted versions.

\begin{definition}
The set of {\em terms} of Higher-Order Datalog is defined as follows:
\begin{itemize}
  \item Every predicate variable (respectively predicate constant) of type $\pi$ is a
        term of type $\pi$; every individual variable (respectively individual constant)
        of type $\iota$ is a term of type $\iota$;
  \item if $\mathsf{E}_1$ is a term of type $\rho \rightarrow \pi$ and
        $\mathsf{E}_2$ a term of type $\rho$ then $(\mathsf{E}_1\ \mathsf{E}_2)$ is a term of type $\pi$.
\end{itemize}
\end{definition}
\begin{definition}
The set of {\em expressions} of Higher-Order Datalog is defined as follows:
\begin{itemize}
\item A term of type $\rho$ is an expression of type $\rho$;
\item if $\mathsf{E}_1$ and $\mathsf{E}_2$ are terms of type $\iota$, then $(\mathsf{E}_1\approx \mathsf{E}_2)$ is an expression of type $o$.
\end{itemize}
\end{definition}
Expressions (respectively terms) that have no variables will often be referred to as {\em ground expressions} (respectively {\em ground terms}). Expressions of type $o$ will often be referred to as {\em atoms}. We will omit parentheses
when no confusion arises. To denote that an expression $\mathsf{E}$ has type $\rho$ we will often write $\mathsf{E}:\rho$.

\begin{definition}\label{definitional_programs}
A {\em clause} (or {\em rule}) of Higher-Order Datalog is a formula
$\mathsf{p}\ \mathsf{V}_1 \cdots \mathsf{V}_n \leftarrow \mathsf{E}_1 \wedge \cdots \wedge \mathsf{E}_m$,
where $\mathsf{p}$ is a predicate constant of type $\rho_1 \rightarrow \cdots \rightarrow\rho_n \rightarrow o$,
$\mathsf{V}_1,\ldots,\mathsf{V}_n$, $n\geq 0$, are argument variables of types $\rho_1,\ldots,\rho_n$ respectively, and $\mathsf{E}_1,\ldots,\mathsf{E}_m$, $m\geq 0$, are atoms. The term $\mathsf{p}\ \mathsf{V}_1 \cdots \mathsf{V}_n$
is called the {\em head} of the clause, the variables $\mathsf{V}_1, \ldots, \mathsf{V}_n$ are the {\em formal parameters} of the
clause and the conjunction $\mathsf{E}_1 \wedge \cdots \wedge \mathsf{E}_m$ is its {\em body}.
A {\em definitional clause} is a clause that additionally satisfies the following
two restrictions:
\begin{enumerate}
\item All the formal parameters are distinct variables (i.e., for all $i,j$ such that $1\leq i,j\leq n$ and $i\neq j$,
      $\mathsf{V}_i \neq \mathsf{V}_j$).

\item The only variables that can appear in the body of the clause are its formal parameters
      and possibly some additional individual variables (variables of type $\iota$).
\end{enumerate}
A {\em definitional program} $\mathsf{P}$ of Higher-Order Datalog is a set of definitional program clauses.
\end{definition}
In the rest of the paper, when we refer to ``clauses'' and ``programs'' we will mean definitional ones.
Notice that for uniformity reasons, the above definition requires that {\em all} formal parameters are distinct,
even the type $\iota$ ones. This is not a real restriction, because two occurrences of the same individual variable
in the head of a clause can be replaced by distinct variables, which are then explicitly equated in the body
of the clause using the constant $\approx$.

\begin{example}
\label{exa_alt_syntax}
Assume that {\tt p} is of type $\iota \rightarrow o$, {\tt q} of type
$\iota \rightarrow \iota \rightarrow o$ and {\tt r} of type
$(\iota \rightarrow o) \rightarrow (\iota \rightarrow o) \rightarrow \iota \rightarrow o$.
The following is a legitimate program of Higher-Order Datalog:
\[
\begin{array}{l}
\mbox{\tt p X $\leftarrow $ (X $\approx$ a)}\\
\mbox{\tt q X Y $\leftarrow $ (X $\approx$ Y)}\\
\mbox{\tt r P Q X $\leftarrow$ (X $\approx$ b) $\wedge$ (P X) $\wedge$ (Q Y)}
\end{array}
\]
Notice that the formal parameters of every clause are distinct.
\end{example}
The above somewhat rigid syntax is quite convenient for formal
purposes. We will use it in the following subsection when discussing
the semantics of Higher-Order Datalog. However, in the rest of the
paper we will relax the above strict notation and write in a more
Prolog-like syntax.
\begin{example}
The program of the previous example will be written in the following simpler
form:
\[
\begin{array}{l}
\mbox{\tt p a.}\\
\mbox{\tt q X X.}\\
\mbox{\tt r P Q b $\leftarrow$ (P b),(Q Y).}
\end{array}
\]
In other words, we will allow some common Prolog conventions, such as the
usual fact syntax, using the comma instead of $\wedge$, allowing
individual constants to appear in the heads of clauses, having multiple occurrences
of the same individual variable in the head of a clause, and using
the full stop to end clauses. Obviously, every program that uses the
Prolog-like syntax can be transformed into the more formal one.
\end{example}
\begin{remark}
The syntax described above, although somewhat close to the traditional syntax
of first-order logic programming, differs from it in one important respect: when defining
predicates, we do not use the common tuple notation but instead we have adopted
the ``{\em currying syntax}'' that is standard in functional programming languages.
For example, in the head of the clause defining the predicate {\tt r} above,
we write {\tt r P Q b} instead of the more common {\tt r(P,Q,b)}. Currying is an important
tool that allows functions (or predicates in our case) to be {\em partially applied}, ie.,
invoked with less arguments than their full arity. As we are going to see in the next sections,
partial applications play an important role in our constructions.
\end{remark}

The {\em Herbrand universe} $U_{\mathsf{P}}$ of a program $\mathsf{P}$ is the set of
constants that appear in $\mathsf{P}$ (if no constant appears in $\mathsf{P}$, we select an
arbitrary one).

In the rest of the paper we will consider fragments of Higher-Order Datalog
based on the {\em order} of predicates that appear in programs:
\begin{definition}
The {\em order} of a type is recursively defined as follows:
\[
\begin{array}{rcl}
\textit{order}(\iota) & = & 0\\
\textit{order}(o)     & = & 0\\
\textit{order}(\rho_1 \rightarrow \cdots \rightarrow \rho_n \rightarrow o) & = & 1+\textit{max}(\{\textit{order}(\rho_i) \mid 1 \leq i \leq n\})
\end{array}
\]
The order of a predicate constant (or variable) is the order of its type.
\end{definition}
\begin{definition}
For all $k\geq 1$, {\em $k$-order Datalog} is the fragment of Higher-Order Datalog
in which all predicate constants have order less than or equal to $k$ and all predicate
variables have order less than or equal to $k-1$.
\end{definition}
\begin{example}
Consider again the program of Example~\ref{exa_alt_syntax}. Predicates {\tt p} and {\tt q} are
ordinary first-order ones. The order of {\tt r} is:
$$\textit{order}((\iota \rightarrow o) \rightarrow (\iota \rightarrow o) \rightarrow \iota \rightarrow o) =
1+\textit{max}(\{\textit{order}(\iota \rightarrow o),\textit{order}(\iota)\}) = 2$$
This program belongs to 2nd-order Datalog because the predicate constants {\tt p}, {\tt q} and {\tt r}
have order  less than or equal to 2 and the predicate variables {\tt P} and {\tt Q} have order 1.
\end{example}

\subsection{The Semantics of Higher-Order Datalog}\label{semantics}
In this subsection we present the semantics of Higher-Order Datalog, which is
based on the ideas initially proposed in~\cite{Wa91a} and subsequently extended and refined
in~\cite{KRW05,CHRW13}. The key idea is to interpret program predicates as monotonic relations.
This ensures that the immediate consequence operator of the program (see Definition~\ref{tp}
later in this subsection), is also monotonic and therefore has a least fixpoint.

We start by defining the semantics of the types of our language. More specifically, we define
simultaneously and recursively the semantics $\mo{\rho}$ of a type $\rho$ and a
corresponding partial order $\aleq[\rho]$ on the elements of $\mo{\rho}$. We adopt the usual
ordering of the truth values $\mathit{false}$ and $\mathit{true}$, i.e. $\mathit{false} \leq \mathit{false}$,
$\mathit{true}\leq \mathit{true}$ and $\mathit{false} \leq \mathit{true}$. Given posets $A$ and
$B$, we write $[A \stackrel{m}{\rightarrow} B]$ to denote the set of all monotonic functions
from $A$ to $B$.
\begin{definition}
Let $\mathsf{P}$ be a program. Then:
\begin{itemize}
  \item $\mo{\iota} = U_{\mathsf{P}}$ and $\aleq[\iota]$ is the trivial partial order
        that relates every element of $U_{\mathsf{P}}$ to itself;
  \item $\mo{o} = \{ \mathit{false}, \mathit{true} \}$ and $\aleq[o]$ is the partial order $\leq$ on truth values;
  \item $\mo{\rho \rightarrow \pi} = [ \mo{\rho} \stackrel{m}{\rightarrow}  \mo{\pi} ]$ and $\aleq[\rho \rightarrow \pi]$
  is the partial order defined as follows: for all $f, g \in \mo{\rho \rightarrow \pi}$,
  $f \aleq[\rho \rightarrow \pi] g$ iff $f(d) \aleq[\pi] g(d)$ for all $d \in \mo{\rho}$.
\end{itemize}
\end{definition}
We now proceed to define Herbrand interpretations and states.
\begin{definition}
A {\em Herbrand interpretation $I$ of a program $\mathsf{P}$} is a function that assigns:
\begin{itemize}
  \item to each individual constant $\mathsf{c}$ that appears in $\mathsf{P}$, the element $I(\mathsf{c}) = \mathsf{c}$;
  \item to each predicate constant $\mathsf{p} : \pi$ that appears in $\mathsf{P}$, an element $I(\mathsf{p}) \in \mo{\pi}$;
\end{itemize}
\end{definition}

\begin{definition}
  A {\em Herbrand state} $s$ of a program $\mathsf{P}$ is a function that assigns to each argument variable $\mathsf{V}$
  of type $\rho$, an element $s(\mathsf{V}) \in \mo{\rho}$.
\end{definition}
In the following, $s[\mathsf{V}_1/d_1,\ldots,\mathsf{V}_n/d_n]$ is used to denote a state that is identical to $s$ the
only difference being that the new state assigns to each $\mathsf{V}_i$ the corresponding value $d_i$.

\begin{definition}
  Let $\mathsf{P}$ be a program, $I$ a Herbrand interpretation, and
  $s$ a Herbrand state of $\mathsf{P}$. Then, the semantics of expressions is
  defined as follows:
\begin{itemize}
  \item $\mwrs{\mathsf{V}}{I}{s} = s(\mathsf{V})$;
  \item $\mwrs{\mathsf{c}}{I}{s} = I(\mathsf{c})$;
  \item $\mwrs{\mathsf{p}}{I}{s} = I(\mathsf{p})$;
  \item $\mwrs{(\mathsf{E}_1\ \mathsf{E}_2)}{I}{s} = \mwrs{\mathsf{E}_1}{I}{s}(\mwrs{\mathsf{E}_2}{I}{s})$;
  \item $\mwrs{(\mathsf{E}_1\approx \mathsf{E}_2)}{I}{s} = true$ if $\mwrs{\mathsf{E}_1}{I}{s}= \mwrs{\mathsf{E}_2}{I}{s}$ and $\mathit{false}$ otherwise.
\end{itemize}
\end{definition}
For ground expressions $\mathsf{E}$ we will often write $\mwrs{\mathsf{E}}{I}{}$ instead
of $\mwrs{\mathsf{E}}{I}{s}$ since in this case the meaning of $\mathsf{E}$ is independent of $s$.
The notion of {\em model} is defined as follows:
\begin{definition}
Let $\mathsf{P}$ be a program and $M$ be a Herbrand interpretation of $\mathsf{P}$.
Then, $M$ is a {\em Herbrand model} of $\mathsf{P}$ iff for every clause
$\mathsf{p}\ \mathsf{V}_1\cdots\mathsf{V}_n \leftarrow \mathsf{E}_1 \wedge \cdots \wedge \mathsf{E}_m$ in $\mathsf{P}$
and for every Herbrand state $s$, if for all $i \in \{1,\ldots,m\}$, $\mwrs{\mathsf{E}_i}{M}{s} = \mathit{true}$ then
$\mwrs{\mathsf{p}\ \mathsf{V}_1\cdots\mathsf{V}_n}{M}{s} = \mathit{true}$.
\end{definition}
\begin{example}
Consider the following program, where {\tt p} is of type $\iota \rightarrow o$ and {\tt q} of type
$(\iota \rightarrow o) \rightarrow o$:
\[
\begin{array}{l}
\mbox{\tt p a.}\\
\mbox{\tt q R $\leftarrow$ (R b).}
\end{array}
\]
It can easily be seen that the Herbrand interpretation that assigns to {\tt p} the relation $\{ {\tt a} \}$ and to {\tt q}
the relation $\{\{{\tt b}\},\{{\tt a},{\tt b}\}\}$, is a model of the program. Notice that the meaning of {\tt q}
is monotonic: since it is true of the relation $\{{\tt b}\}$, it has to also be true of the relation $\{{\tt a},{\tt b}\}$
(which is a superset of $\{{\tt b}\}$). Actually, the interpretation we just described is the {\em minimum model} of the
program, a notion that will be discussed shortly.
\end{example}

We denote the set of Herbrand interpretations of a program $\mathsf{P}$
with ${\cal I}_\mathsf{P}$, and define a partial order on ${\cal I}_\mathsf{P}$ as
follows: for all $I, J \in {\cal I}_\mathsf{P}$, $I \aleq[{\cal I}_\mathsf{P}] J$
iff for every predicate constant $\mathsf{p} : \pi$ that appears in $\mathsf{P}$, $I(\mathsf{p}) \aleq[\pi] J(\mathsf{p})$.
It is easy to prove that $({\cal I}_\mathsf{P},\aleq[{\cal I}_\mathsf{P}])$ is a complete lattice;
we denote by $\bigsqcup$ the least upper bound operation and by
$\perp_{{\cal I}_{\mathsf{P}}}$ the least element of the lattice, with respect to
$\aleq[{\cal I}_\mathsf{P}]$. Intuitively, $\perp_{{\cal I}_{\mathsf{P}}}$ assigns to every
program predicate in $\mathsf{P}$ the empty relation.

We can now define the {\em immediate consequence operator} for
Higher-Order Datalog programs, which generalizes the corresponding operator for classical
Datalog~\cite{lloyd}.
\begin{definition}\label{tp}
  Let $\mathsf{P}$ be a program. The mapping $T_\mathsf{P} : {\cal I}_\mathsf{P} \rightarrow {\cal I}_\mathsf{P}$
  is called the {\em immediate consequence operator for $\mathsf{P}$} and is defined for every predicate constant
  $\mathsf{p} : \rho_1 \rightarrow \cdots \rightarrow \rho_n \rightarrow o$ and $d_i \in \mo{\rho_i}$ as:
\[T_\mathsf{P}(I)(\mathsf{p})\ d_1\cdots d_n =
\begin{cases}
  \mathit{true},  & \mbox{if there exists a clause $\mathsf{p}\ \mathsf{V}_1\cdots\mathsf{V}_n \leftarrow \mathsf{E}_1 \wedge \cdots \wedge \mathsf{E}_m$ in $\mathsf{P}$ and}\\
                 & \mbox{a Herbrand state $s$, such that  $\mwrs{\mathsf{E}_i}{I}{s[\mathsf{V}_1/d_1, \ldots, \mathsf{V}_n/d_n]} = \mathit{true}$}\\
                 & \mbox{for all $i \in \{1, \ldots, m\}$} \\
  \mathit{false}, & \mbox{otherwise.}
\end{cases}
\]
\end{definition}

Define now the following sequence of interpretations:
\[
\begin{array}{lll}
T_{\mathsf{P}} \uparrow 0  & = & \perp_{{\cal I}_{\mathsf{P}}}\\
T_{\mathsf{P}} \uparrow (n+1)  & = & T_{\mathsf{P}}(T_{\mathsf{P}} \uparrow n)\\
T_{\mathsf{P}} \uparrow \omega & = & \bigsqcup \{ T_{\mathsf{P}} \uparrow n \mid n<\omega\}
\end{array}
\]
We then have the following theorem (see~\cite{Wa91a,KRW05,CHRW13} for more details):
\begin{theorem}
Let $\mathsf{P}$ be a program and let $M_{\mathsf{P}} = T_{\mathsf{P}}\uparrow \omega$. Then, $M_{\mathsf{P}}$
is the least Herbrand model of $\mathsf{P}$ and the least fixpoint of $T_\mathsf{P}$
(with respect to the ordering relation $\aleq[{\cal I}_\mathsf{P}]$).
\end{theorem}

\section{Decision Problems, Logic Programming, and Complexity Classes}
\label{complexity_basics}
In this section we initiate our investigation regarding the expressive power
of Higher-Order Datalog. Our development is based on well-known ideas relating
logic programming languages with complexity theory (see for example~\cite{DEGV01}
for an introduction of the main concepts).

Let $\Sigma$ be an alphabet. Without loss of generality, we fix $\Sigma=\{a,b\}$.
Our goal is to demonstrate that sublanguages of Higher-Order Datalog correspond to interesting
complexity classes over $\Sigma$. We first need to specify how strings over $\Sigma$
can be encoded in our setting. For this purpose, we use a ternary predicate
{\tt input} which encodes in an {\em ordered} manner the input string. For example,
to represent the string {\tt abba} we use the four facts:
\[
\begin{array}{l}
\mbox{\tt input 0 a 1.}\\
\mbox{\tt input 1 b 2.}\\
\mbox{\tt input 2 b 3.}\\
\mbox{\tt input 3 a end.}
\end{array}
\]
More generally, an input of length $n>0$ over $\Sigma$ can be encoded with $n$ facts
of the above form. Moreover, for input of length $n=0$, namely for the empty string,
we use:
\[
\begin{array}{l}
\mbox{\tt input 0 empty  end.}
\end{array}
\]
where {\tt empty} is a constant that denotes the empty string.

Given string $w \in \Sigma^*$, we will write ${\cal D}_w$ to denote the set of
facts that represent $w$ through the {\tt input} relation. This encoding of input
strings is usually referred as the {\em ordered database assumption}.

We will assume that every program defines a propositional {\tt accept} predicate which,
intuitively, signals whether a particular input string is accepted by our program. We have
the following two definitions:
\begin{definition}
Let $\Sigma$ be an alphabet. We will say that a Higher-Order Datalog program $\mathsf{P}$ {\em decides}
a language $L \subseteq \Sigma^*$ if for any $w \in \Sigma^*$, $w \in L$ iff {\tt accept}
is true in the minimum Herbrand model of $\mathsf{P}\cup {\cal D}_w$.
\end{definition}
\begin{definition}
We will say that a set ${\cal Q}$ of Higher-Order Datalog programs captures the
complexity class ${\cal C}$, if the set of languages decided by the programs
in ${\cal Q}$ coincides with ${\cal C}$.
\end{definition}
Assuming the above representation of input strings through the {\tt input} relation,
the following classical result has been obtained in many different contexts~\cite{Pap85,Gra92,Var82,Imm86,Lei89}:
\begin{theorem}\label{Datalog_captures_PTIME}
The set of first-order Datalog programs captures $\mathsf{PTIME}$.
\end{theorem}
We give a detailed proof of this theorem in~\ref{PTIME} by refining the
expositions given in~\cite{Pap85} and~\cite{DEGV01}. Although this is a well-known
result, the reader is advised to first look through this proof before attempting
to read the more involved ones in the rest of the paper. Actually, several ideas
and predicates  defined in~\ref{PTIME} are needed to define the predicates
for the higher-order case.

We now proceed to examine the expressive power of Higher-Order Datalog.
We will need the following family of functions:
\[
\begin{array}{rcl}
\expk{0}(x)      & = & x\\
\expk{n+1}(x)    & = & 2^{\expk{n}(x)}
\end{array}
\]
For all $k\geq 0$, the complexity class $k-\mathsf{EXPTIME}$ is defined as follows:
$$k-\mathsf{EXPTIME}  =  \bigcup_{r\in \mathbb{N}} \mathsf{TIME}(\expk{k}(n^r))$$
Notice that $0-\mathsf{EXPTIME}$ coincides with $\mathsf{PTIME}$.
The following theorem, which we will establish, is an extension of Theorem~\ref{Datalog_captures_PTIME} to the
case of Higher-Order Datalog:
\begin{theorem}\label{main_theorem}
For every $k\geq 1$, the set of $k$-order Datalog programs captures $(k-1)$-$\mathsf{EXPTIME}$.
\end{theorem}
Obviously, for $k=1$ the above theorem gives as a special case Theorem~\ref{Datalog_captures_PTIME}.
The detailed proof of Theorem~\ref{main_theorem} for the cases $k\geq 2$, is developed in the next section.

\section{Higher-Order Datalog and Exponential Time Bounded Turing Machines}
\label{equivalence}

In order to establish Theorem~\ref{main_theorem}, we prove two lemmas. The first one
shows that every language decided by a $k$-order Datalog program can also be decided
by a $(k-1)$-exponential time bounded Turing machine. More specifically:
\begin{lemma}\label{first_direction}
Let $\mathsf{P}$ be a $k$-order Datalog program, $k\geq 2$, that decides
a language $L$. Then, there exists a Turing machine that decides $L$
in time $O(\expk{k-1}(n^q))$, where $n$ is the length of the input string
and $q$ is a constant that depends only on $\mathsf{P}$.
\end{lemma}
The proof of the above lemma is given in~\ref{appendix_b}, and is based on
calculating the time-complexity of the bottom-up proof procedure for
Higher-Order Datalog.

We now demonstrate the following lemma, which is the converse of
Lemma~\ref{first_direction}:

\begin{lemma}
Let $M$ be a deterministic Turing machine that decides a language $L$ in
$(k-1)-\mathsf{EXPTIME}$, $k\geq 2$. Then, there exists a $k$-order Datalog
program $\mathsf{P}$ that decides $L$.
\end{lemma}

The next three subsections establish the proof of the above lemma. The key idea is to construct
a $k$-order Datalog program that simulates the $(k-1)$-exponential-time-bounded Turing
machine $M$. In order to achieve this, we must use the power of higher-order relations
to represent ``large numbers'' that count the execution steps of the machine. As it turns out,
by increasing the order of the programs that we use, we can increase the range of representable numbers.

Assume that $M$ decides $L$ in time $O(\expk{k-1}(n^q))$. Then there exists an integer constant $d$,
such that for every input $w$ of length $n \geq 2$, $M$ terminates after at most $\expk{k-1}(n^d)-1$ steps.
The simulation that we will present, produces the correct answer for all
inputs of length at least 2 by simulating $\expk{k-1}(n^d)-1$ steps of the Turing machine $M$.
Similarly to the first-order case (see~\ref{PTIME}), for the special cases of strings of length 0
or 1 that belong to $L$, the correct answer is produced directly by appropriate rules.

In Subsection~\ref{case_two} we demonstrate that for any $d>0$, there exists a
second-order Datalog program which, given any {\tt input} relation of size $n$, can
represent all natural numbers up to $2^{n^d}-1$. In Subsection~\ref{case_greater_than_two}
we show that using $k$-order Datalog, $k>2$, we can represent numbers up to $\expk{k-1}(n^d) - 1$.
Notice that the case $k=2$ has some differences from the case where $k>2$, and that's
why we devote two different subsections to the two cases. The differences
are due to the fact that the simulation for $k=2$ uses tuples in order to represent
numbers, while the simulation for $k>2$ uses higher-order predicates for the same
purpose.

Finally, in Subsection~\ref{Turing_Simulation} we provide the actual simulation of the
$(k-1)$-exponential-time-bounded Turing machine $M$ by the $k$-order Datalog
program.

\subsection{The Second-Order Case}\label{case_two}
In this subsection we demonstrate that we can use second-order relations
to represent numbers up to $2^{n^d} - 1$. We use a technique similar to the
one introduced in~\cite{Jones}: a number in the range $0,\ldots,2^{n^d}-1$
can be represented by a function $f:\{0,\ldots,n^d-1\} \rightarrow \{0,1\}$.
Such a function is equivalent to a string of $n^d$ binary digits. We assume
that $f(0)$ is the rightmost bit of the number and $f(n^d - 1)$ the
leftmost one. Such a string can represent any number in the required range.

Second-order Datalog can implement a function such as the above using a
binary predicate {\tt p $\overline{\tt X}$ V}, where  $\overline{\tt X}={\tt X}_1\cdots {\tt X}_d$
is a $d$-tuple (alternatively, $d$ consecutive arguments) that can represent all numbers
in the range $0$ to $n^d-1$ (please see~\ref{PTIME}), and {\tt V} is a variable
that can receive either the constant {\tt low} or the constant {\tt high}
(corresponding to 0 and 1 respectively).

The main predicates that are defined in this subsection are:
${\tt zero}_1$, ${\tt last}_1$, {\tt is\_zero}$_1$, ${\tt is\_non\_zero}_1$,
${\tt is\_last}_1$, ${\tt non\_last}_1$, {\tt pred}$_1$, {\tt succ}$_1$,
{\tt equal}$_1$, and {\tt less\_than}$_1$. We explain the purpose of each
one of them, just before we define it. The subscript 1 in all the above predicates,
denotes that we are now using first-order relations in order to represent
our numbers (in~\ref{PTIME} we represented smaller numbers using $d$-tuples).
Notice that we will also use some additional auxiliary predicates in our definitions
as-well-as some predicates from the first-order case defined in~\ref{PTIME}
(namely, {\tt tuple\_zero}, {\tt tuple\_last}, {\tt tuple\_pred}).

We start by defining the predicates ${\tt zero}_1$ and ${\tt last}_1$ that represent
the first and last numbers of the range (namely 0 and $2^{n^d}-1$ respectively).
{\tt
\begin{center}
\begin{tabular}{l}
zero$_1$ $\overline{\tt X}$ low.\\
last$_1$ $\overline{\tt X}$ high.
\end{tabular}
\end{center}
}
Notice that {\tt zero}$_1$ returns in its second argument the value {\tt low} for all
values of its first argument (and similarly for {\tt last}$_1$ and {\tt high}).
We now define a predicate {\tt is\_zero}$_1$ that checks if its argument
is equal to the function {\tt zero}$_1$. The auxiliary predicate {\tt (all\_to\_right$_1$ V N $\overline{\tt X}$)}
checks if all the bits of {\tt N} starting from the position indicated by $\overline{\tt X}$ until
the right end of {\tt N}, have the value {\tt V} ({\tt low} in our case).
{\tt
\begin{center}
\begin{tabular}{lll}
is\_zero$_1$ N & $\leftarrow$ &  (tuple\_last $\overline{\tt X}$),(all\_to\_right$_1$ low N $\overline{\tt X}$). \\
\\
all\_to\_right$_1$ V N $\overline{\tt X}$ & $\leftarrow$ & (tuple\_zero $\overline{\tt X}$),(N $\overline{\tt X}$ V).\\
all\_to\_right$_1$ V N $\overline{\tt X}$ & $\leftarrow$ & (tuple\_pred $\overline{\tt X}$ $\overline{\tt Y}$),(N $\overline{\tt X}$ V),(all\_to\_right$_{1}$ V N $\overline{\tt Y}$).
\end{tabular}
\end{center}
}
Similarly we define the predicate ${\tt is\_non\_zero}_1$ that succeeds if its argument
is not equal to the function ${\tt zero}_1$. The predicate {\tt (exists\_to\_right$_1$ V N $\overline{\tt X}$)}
checks if there exists a bit of {\tt N}, starting from the position indicated by $\overline{\tt X}$ until
the right end of {\tt N}, that has the value {\tt V} ({\tt high} in our case).
{\tt
\begin{center}
\begin{tabular}{lll}
non\_zero$_1$ N & $\leftarrow$ & (tuple\_last $\overline{\tt X}$),(exists\_to\_right$_1$ high N $\overline{\tt X}$). \\
\\
exists\_to\_right$_1$ V N $\overline{\tt X}$ & $\leftarrow$ & (N $\overline{\tt X}$ V).\\
exists\_to\_right$_1$ V N $\overline{\tt X}$ & $\leftarrow$ & (tuple\_pred $\overline{\tt X}$ $\overline{\tt Y}$),(exists\_to\_right$_1$ V N $\overline{\tt Y}$).\\
\end{tabular}
\end{center}
}
Symmetrically, we can define the predicates ${\tt is\_last}_1$ and ${\tt non\_last}_1$, as follows:
{\tt
\begin{center}
\begin{tabular}{lll}
is\_last$_1$ N & $\leftarrow$ &  (tuple\_last $\overline{\tt X}$),(all\_to\_right$_1$ high N $\overline{\tt X}$).\\\\

non\_last$_1$ N & $\leftarrow$ & (tuple\_last $\overline{\tt X}$),(exists\_to\_right$_1$ low N $\overline{\tt X}$).
\end{tabular}
\end{center}
}
Next we define ${\tt pred}_1$ to capture the notion of the predecessor of
a number. This is one of the cases where partial application and currying
(see Remark in Section~\ref{syntax}) plays an important role in our encoding of big numbers.
More specifically, the predicate ${\tt pred}_1$ is different than the ${\tt tuple\_pred}$
predicate (see~\ref{PTIME}), in the sense that it does not check if one number is the predecessor of another
number; instead, if {\tt N} is the representation of a number $n$ then the partially
applied expression $({\tt pred}_1 \ {\tt N})$ is the representation of the predecessor of {\tt N}.
The idea is that the predecessor of {\tt N}, namely {\tt (pred$_1$ N)}, is a number
whose binary representation has at position $\overline{\tt X}$ either: (i) the same binary digit
as {\tt N} if there exists some bit of {\tt N} that is on the right of position $\overline{\tt X}$
that has the value {\tt high}, or (ii) the inverse binary digit of that of {\tt N} at position $\overline{\tt X}$,
if all the bits of {\tt N} that are on the right of $\overline{\tt X}$ have the value {\tt low}.
{\tt
\begin{center}
\begin{tabular}{lll}
pred$_{1}$ N $\overline{\tt X}$ V & $\leftarrow$ & (tuple\_zero $\overline{\tt X}$),(non\_zero$_{1}$ N),\\
                                  &              & (N $\overline{\tt X}$ V1),(invert V1 V).\\
pred$_{1}$ N $\overline{\tt X}$ V & $\leftarrow$ & (non\_zero$_{1}$ N),(tuple\_pred $\overline{\tt X}$ $\overline{\tt Y}$),\\
                                  &              & (exists\_to\_right$_{1}$ high N $\overline{\tt Y}$),(N $\overline{\tt X}$ V).\\
pred$_{1}$ N $\overline{\tt X}$ V & $\leftarrow$ & (non\_zero$_{1}$ N),(tuple\_pred $\overline{\tt X}$ $\overline{\tt Y}$),\\
                                  &              & (all\_to\_right$_{1}$ low N $\overline{\tt Y}$),\\
                                  &              & (N $\overline{\tt X}$ V1),(invert V1 V).\\\\

invert low high.                        &              &\\
invert high low.                        &              &\\
\end{tabular}
\end{center}
}
Symmetrically, way define {\tt succ}$_1$ which gives the successor of a given number:
{\tt
\begin{center}
\begin{tabular}{lll}
succ$_{1}$ N $\overline{\tt X}$ V & $\leftarrow$ & (tuple\_zero $\overline{\tt X}$),(non\_last$_{1}$ N),\\
                                  &              & (N $\overline{\tt X}$ V1),(invert V1 V).\\
succ$_{1}$ N $\overline{\tt X}$ V & $\leftarrow$ & (non\_last$_{1}$ N),(tuple\_pred $\overline{\tt X}$ $\overline{\tt Y}$),\\
                                  &              & (exists\_to\_right$_{1}$ low N $\overline{\tt Y}$),(N $\overline{\tt X}$ V).\\
succ$_{1}$ N $\overline{\tt X}$ V & $\leftarrow$ & (non\_last$_{1}$ N),(tuple\_pred $\overline{\tt X}$ $\overline{\tt Y}$),\\
                                  &              & (all\_to\_right$_{1}$ high N $\overline{\tt Y}$),\\
                                  &              & (N $\overline{\tt X}$ V1),(invert V1 V).
\end{tabular}
\end{center}
}
We will also need the equality of two numbers {\tt N} and {\tt M}. We compare them bit by bit,
starting from the leftmost possible position and moving to the left.
{\tt
\begin{center}
\begin{tabular}{lll}
equal$_{1}$ N M                                  & $\leftarrow$ & (tuple\_last $\overline{\tt X}$),(equal\_test$_{1}$ N M $\overline{\tt X}$).\\\\
equal\_test$_{1}$ N M $\overline{\tt X}$         & $\leftarrow$ & (tuple\_zero $\overline{\tt X}$),(N $\overline{\tt X}$ V),(M $\overline{\tt X}$ V).\\
equal\_test$_{1}$ N M $\overline{\tt X}$         & $\leftarrow$ & (tuple\_pred $\overline{\tt X}$ $\overline{\tt Y}$),(N $\overline{\tt X}$ V),(M $\overline{\tt X}$ V),\\
                                                 &              & (equal\_test$_{1}$ N M $\overline{\tt Y}$).
\end{tabular}
\end{center}
}
Finally, we will need the ``less-than'' relation, defined as follows:
{\tt
\begin{center}
\begin{tabular}{lll}
less\_than$_{1}$ N M         & $\leftarrow$ & (is\_zero$_{1}$ N),(non\_zero$_{1}$ M).\\
less\_than$_{1}$ N M         & $\leftarrow$ & (non\_zero$_{1}$ N),(non\_zero$_{1}$ M),\\
                               &              & (less\_than$_{1}$ (pred$_{1}$ N) (pred$_{1}$ M)).
\end{tabular}
\end{center}
}
In order to represent even larger numbers, we need to extend the above predicates to higher orders. This
requires certain modifications to the predicate definitions, as the following subsection demonstrates.

\subsection{Extending to Arbitrary Orders}\label{case_greater_than_two}
To define numbers larger than $2^{n^d}$, we need to generalize the ideas of
the previous subsection. For example, a number in the range $0,\ldots,2^{2^{n^d}}-1$
can be represented by a function $f:\{0,\ldots,2^{n^d}-1\} \rightarrow \{0,1\}$.
In other words, to define the numbers and the operations at level $k+1$, for $k\geq 2$,
we need to use the numbers and the operations of level $k$. The definitions
we give below, have certain differences from the second-order ones given in the previous
subsection. This is due to the fact that the second-order predicates, use the
tuple-based predicates that are defined in~\ref{PTIME} (while the ones we define
below, do not). We start with ${\tt zero}_{k+1}$ and ${\tt last}_{k+1}$. Notice that
the parameter {\tt X} is now a relation (and not a tuple as in the case of {\tt zero}$_1$
and {\tt last}$_1$).
{\tt
\begin{center}
\begin{tabular}{l}
zero$_{k+1}$ X low.\\
last$_{k+1}$ X high.
\end{tabular}
\end{center}
}
We now define {\tt is\_zero}$_{k+1}$ which succeeds if its argument is equal
to {\tt zero}$_{k+1}$:
{\tt
\begin{center}
\begin{tabular}{lll}
is\_zero$_{k+1}$ N & $\leftarrow$ &  (all\_to\_right$_{k+1}$ low N last$_k$). \\
\\
all\_to\_right$_{k+1}$ V N X & $\leftarrow$ & (is\_zero$_k$ X),(N X V).\\
all\_to\_right$_{k+1}$ V N X & $\leftarrow$ & (non\_zero$_k$ X),(N X V),\\
                             &              & (all\_to\_right$_{k+1}$ V N (pred$_k$ X)).
\end{tabular}
\end{center}
}
Similarly we define the predicate ${\tt non\_zero}_{k+1}$ that succeeds if its argument
is not equal to ${\tt zero}_{k+1}$:
{\tt
\begin{center}
\begin{tabular}{lll}
non\_zero$_{k+1}$ N & $\leftarrow$ & (exists\_to\_right$_{k+1}$ high N last$_k$). \\
\\
exists\_to\_right$_{k+1}$ V N X & $\leftarrow$ & (N X V).\\
exists\_to\_right$_{k+1}$ V N X & $\leftarrow$ & (non\_zero$_{k}$ X),\\
                                &              & (exists\_to\_right$_{k+1}$ V N (pred$_k$ X)).
\end{tabular}
\end{center}
}
Symmetrically, we define the predicates ${\tt is\_last}_{k+1}$ and ${\tt non\_last}_{k+1}$, as follows:
{\tt
\begin{center}
\begin{tabular}{lll}
is\_last$_{k+1}$  N & $\leftarrow$ & (all\_to\_right$_{k+1}$ high N last$_{k}$).\\\\

non\_last$_{k+1}$ N & $\leftarrow$ & (exists\_to\_right$_{k+1}$ low N last$_{k}$).
\end{tabular}
\end{center}
}
Using the above predicates we can now define {\tt pred}$_{k+1}$ as follows:
{\tt
\begin{center}
\begin{tabular}{lll}
pred$_{k+1}$ N X V & $\leftarrow$ & (is\_zero$_k$ X),(non\_zero$_{k+1}$ N),\\
                   &              & (N X V1),(invert V1 V).\\
pred$_{k+1}$ N X V & $\leftarrow$ & (non\_zero$_{k}$ X),\\
                   &              & (exists\_to\_right$_{k+1}$ high N (pred$_k$ X)),(N X V).\\
pred$_{k+1}$ N X V & $\leftarrow$ & (non\_zero$_{k}$ X),(non\_zero$_{k+1}$ N),\\
                   &              & (all\_to\_right$_{k+1}$ low N (pred$_k$ X)),\\
                   &              & (N X V1),(invert V1 V).
\end{tabular}
\end{center}
}
In a symmetric way we define the predicate {\tt succ}$_{k+1}$ as follows:
{\tt
\begin{center}
\begin{tabular}{lll}
succ$_{k+1}$ N X V & $\leftarrow$ & (is\_zero$_k$ X),(non\_last$_{k+1}$ N),\\
                   &              & (N X V1),(invert V1 V).\\
succ$_{k+1}$ N X V & $\leftarrow$ & (non\_zero$_{k}$ X),\\
                   &              & (exists\_to\_right$_{k+1}$ low N (pred$_k$ X)),(N X V).\\
succ$_{k+1}$ N X V & $\leftarrow$ & (non\_zero$_{k}$ X),(non\_zero$_{k+1}$ N),\\
                   &              & (all\_to\_right$_{k+1}$ high N (pred$_k$ X)),\\
                   &              & (N X V1),(invert V1 V).
\end{tabular}
\end{center}
}
Equality of two numbers is defined as follows:
{\tt
\begin{center}
\begin{tabular}{lll}
equal$_{k+1}$ N M                 & $\leftarrow$ & (equal\_test$_{k+1}$ N M last$_k$).\\\\
equal\_test$_{k+1}$ N M X         & $\leftarrow$ & (is\_zero$_{k}$ X),(N X V),(M X V).\\
equal\_test$_{k+1}$ N M X         & $\leftarrow$ & (non\_zero$_{k}$ X),(N X V),(M X V),\\
                                  &              & (equal\_test$_{k+1}$ N M (pred$_k$ X)).
\end{tabular}
\end{center}
}
Finally, {\tt less\_than$_{k+1}$} can be defined in an identical way as
in the previous subsection.

\subsection{Simulating Turing Machines with Higher-Order Datalog}\label{Turing_Simulation}
In this section we demonstrate how we can use $k$-order Datalog to simulate
$(k-1)$-exponential-time-bounded Turing machines. The simulation using second-order
programs does not differ from the one that uses programs of order greater than or equal
to three, except for a minor difference described below.

In order to define the initialization rules for the Turing machine, we will need
a predicate that transforms the numbers $0,\ldots,n-1$ that appear in the {\tt input}
relation, to order-$k$ representation of numbers. Recall that such a number is a function
from order-$(k-1)$ numbers to the values {\tt low} and {\tt high}:
{\tt
\begin{center}
\begin{tabular}{l}
base\_to\_higher$_k$ 0 X low. \\
base\_to\_higher$_k$ M X V  \,\, $\leftarrow$ \,\,  (input J $\sigma$ M),(succ$_k$ (base\_to\_higher$_k$ J) X V).
\end{tabular}
\end{center}
}
When $k=2$ the only required change in the above predicate is to replace all the
occurrences of {\tt X} by $\overline{\tt X}$.

The simulation of the exponential-time bounded Turing machine is presented below.
Recall (see~\ref{PTIME}) that we assume that in the beginning of its operation,
the first $n$ squares of the tape hold the input, the rest of the squares hold the
empty character ``\textvisiblespace'' and the machine starts operating from its initial
state denoted by $s_0$. If the Turing machine accepts the input then it goes into the
special state called {\tt yes} and remains in this state forever.

One new feature (with respect to the first-order case in~\ref{PTIME}),
is the representation of the {\tt cursor} predicate. In the first-order case we used
{\tt cursor $\overline{\tt T}$ $\overline{\tt X}$} to mean that at time-point $\overline{\tt T}$
the cursor is located at position $\overline{\tt X}$. In the higher-order case the expression
{\tt (cursor T)} is a number that denotes the position of the cursor, ie., it is a function
from positions to the values {\tt low} and {\tt high}. We start with the initialization rules:

{\tt
\begin{center}
\begin{tabular}{lll}
symbol$_\sigma$ T X  & $\leftarrow$ & (is\_zero$_k$ T),(input Y $\sigma$ W),\\
&        & equal$_k$ (base\_to\_higher$_k$ Y) X.\\
symbol$_{\textvisiblespace}$ T X  & $\leftarrow$ & (is\_zero$_k$ T),(base\_last Y),\\
&        & (less\_than$_{k}$ (base\_to\_higher$_k$ Y) X).\\
state$_{s_0}$ T      & $\leftarrow$ & (is\_zero$_k$ T).\\
cursor T I low       & $\leftarrow$ & (is\_zero$_k$ T).
\end{tabular}
\end{center}
}

We now define the transition rules of the Turing machine. For each transition
rule we generate rules for ${\tt state}_s$, ${\tt symbol}_{\sigma}$ and ${\tt cursor}$.
The transition: ``if the head is in symbol $\sigma$ and in state $s$ then
write symbol $\sigma'$ and go to state $s'$'', generates the following:
{\tt
\begin{center}
\begin{tabular}{lll}
symbol$_{\sigma'}$ T X & $\leftarrow$ &(non\_zero$_k$ T),(equal$_k$ X (cursor (pred$_k$ T))),\\
                       &              &(state$_s$ (pred$_k$ T)),\\
                       &              & (symbol$_\sigma$ (pred$_k$ T) (cursor (pred$_k$ T))).\\
state$_{s'}$ T         & $\leftarrow$ &(non\_zero$_k$ T),(state$_s$ (pred$_k$ T)),\\
                       &              &(symbol$_\sigma$ (pred$_k$ T) (cursor  (pred$_k$ T))). \\
cursor T I V           & $\leftarrow$ &(non\_zero$_k$ T),(state$_s$ (pred$_k$ T)),\\
                       &              &(symbol$_\sigma$ (pred$_k$ T) (cursor  (pred$_k$ T))),\\
                       &              &(cursor (pred$_k$ T) I V).
\end{tabular}
\end{center}
}
We continue with the transition: ``if the head is in symbol $\sigma$ and in state $s$ then
go to state $s'$ and move the head right'', which generates the following:
{\tt
\begin{center}
\begin{tabular}{lll}
symbol$_{\sigma}$ T X  & $\leftarrow$ &(non\_zero$_k$ T),(equal$_k$ X (cursor (pred$_k$ T))),\\
                       &              &(state$_s$ (pred$_k$ T)),\\
                       &              &(symbol$_\sigma$ (pred$_k$ T) (cursor (pred$_k$ T))).\\
state$_{s'}$ T         & $\leftarrow$ &(non\_zero$_k$ T),(state$_s$ (pred$_k$ T)),\\
                       &              &(symbol$_\sigma$ (pred$_k$ T) (cursor  (pred$_k$ T))). \\
cursor T I V           & $\leftarrow$ &(non\_zero$_k$ T),(state$_s$ (pred$_k$ T)),\\
                       &              &(symbol$_\sigma$ (pred$_k$ T) (cursor  (pred$_k$ T))),\\
                       &              &((succ$_k$ (cursor (pred$_k$ T))) I V).
\end{tabular}
\end{center}
}
We also have the transition: ``if the head is in symbol $\sigma$ and in state $s$ then
go to state $s'$ and move the head left'', which generates the following:
{\tt
\begin{center}
\begin{tabular}{lll}
symbol$_{\sigma}$ T X  & $\leftarrow$ &(non\_zero$_k$ T),(equal$_k$ X (cursor (pred$_k$ T))),\\
                       &              &(state$_s$ (pred$_k$ T)),\\
                       &              &(symbol$_\sigma$ (pred$_k$ T) (cursor (pred$_k$ T))).\\
state$_{s'}$ T         & $\leftarrow$ &(non\_zero$_k$ T),(state$_s$ (pred$_k$ T)),\\
                       &              &(symbol$_\sigma$ (pred$_k$ T) (cursor  (pred$_k$ T))). \\
cursor T I V           & $\leftarrow$ &(non\_zero$_k$ T),(state$_s$ (pred$_k$ T)),\\
                       &              &(symbol$_\sigma$ (pred$_k$ T) (cursor (pred$_k$ T))),\\
                       &              &((pred$_k$ (cursor (pred$_k$ T))) I V).
\end{tabular}
\end{center}
}
The inertia rules are the following:
{\tt
\begin{center}
\begin{tabular}{lll}
symbol$_{\sigma}$ T X  & $\leftarrow$ &(less\_than$_k$ X (cursor (pred$_k$ T))),(symbol$_\sigma$ (pred$_k$ T) X).\\
symbol$_{\sigma}$ T X  & $\leftarrow$ &(less\_than$_k$ (cursor (pred$_k$ T)) X),(symbol$_\sigma$ (pred$_k$ T) X).
\end{tabular}
\end{center}
}
Finally, we have the following rule that concerns acceptance:
{\tt
\begin{center}
\begin{tabular}{lll}
accept                      & $\leftarrow$ &(state$_{\tt yes}$ last$_k$).
\end{tabular}
\end{center}
}

When $k=2$ the only required change in the clauses given above, is to replace all the
occurrences of {\tt I} in the definition of {\tt cursor} by $\overline{\tt I}$.

\section{Future Work}\label{future_work}
Theorem~\ref{main_theorem} presents a striking analogy with Theorem
7.17 of~\cite{Jones} where it is shown that read-only functional programs of
order $k\geq 2$ capture $(k-1)$-$\mathsf{EXPTIME}$, and first-order such programs
capture $\mathsf{PTIME}$. The functional language that Jones uses has no direct
relationship with Datalog, and this makes the analogy even more interesting.

As a possible direction for future research, we would like to investigate whether
the complexity results obtained in~\cite{Jones} regarding {\em tail-recursive}
read-only functional programs, can extend to the case of Higher-Order Datalog.
A starting point for this would be to first characterize what tail-recursion
means in the context of Higher-Order Datalog. An additional topic for future
research would be to investigate the complexity-theoretic benefits of adding
negation to Higher-Order Datalog. Such an investigation can be based on the recent
proposal for the well-founded semantics of higher-order logic programs~\cite{CRS18}.

\section*{Acknowledgements}
The research of the first author was supported by the Software and Knowledge Engineering Laboratory (SKEL)
of the Institute of Informatics and Telecommunications of NCSR ``Demokritos''.

\ifarxiv
\vspace{-0.35cm}
\fi

\bibliographystyle{acmtrans}
\bibliography{iclp19} 

   \else

   \fi
\fi

\end{document}